\documentclass[twocolumn, preprintnumbers,amsmath,amssymb,balancelastpage]{revtex4}
\pdfoutput=1

\usepackage{putexPRL}
\usepackage{graphicx}
\usepackage{dsfont}
\usepackage{amsmath}
\usepackage{array}
\usepackage{epstopdf}
\usepackage{enumerate}
\usepackage{youngtab}
\usepackage{tensor}
\usepackage{braket}
\usepackage[normalem]{ulem}
\usepackage{slashed}
\usepackage[aligntableaux=center]{ytableau}
\usepackage[utf8]{inputenc}
\usepackage[
      colorlinks=true,
      linkcolor=blue,
      urlcolor=blue,
      filecolor=black,
      citecolor=red,
      ]{hyperref}
\usepackage{braket}
\usepackage{mathtools}
\usepackage{rotating}
\usepackage{tabularx}
\newcolumntype{Y}{>{\centering\arraybackslash}X}
\newcolumntype{C}[1]{>{\centering\arraybackslash}p{#1}}
\usepackage{todonotes}
\usepackage{xcolor}
\definecolor{LightCyan}{rgb}{0.7,1,1}
\definecolor{Gray}{gray}{0.9}
\usepackage{xspace}

\newcommand {\be} {\begin {equation}}
\newcommand {\ee} {\end {equation}}

\newcommand {\bes} {\begin {equation*}}
\newcommand {\ees} {\end {equation*}}

\newcommand{\es}[2] {\begin{equation} \label{#1} \begin{split} #2 \end{split} \end{equation}}

\newcommand{\cG}{{\mathcal G}}

\newcommand{\cM}{{\mathcal M}}

\newcommand{\beq}{\begin{equation}}
\newcommand{\eeq}{\end{equation}}

\def\ie{\begin{equation}\begin{aligned}}
\def\fe{\end{aligned}\end{equation}}

\def\<{\langle}
\def\>{\rangle}

\def\beg{\begin{equation}\begin{gathered}}
\def\eeg{\end{gathered}\end{equation}}

\def\bea{\begin{equation}\begin{aligned}}
\def\eea{\end{aligned}\end{equation}}

\begin{document}

\title{Pure anti-de Sitter supergravity and the conformal bootstrap}

\author{Luis F. Alday}
\affiliation{Mathematical Institute, University of Oxford,
Woodstock Road, Oxford, OX2 6GG, UK}
\author{Shai M.~Chester}         
\affiliation{Department of Particle Physics and Astrophysics, Weizmann Institute of Science, Rehovot, Israel}

\begin{abstract}
We consider graviton scattering in maximal supergravity on Anti-de Sitter space (AdS) in $d+1$ dimensions for $d=3,4,\text{and $6$}$ with no extra compact spacetime factor. Holography suggests that this theory is dual to an exotic maximally supersymmetric conformal field theory (CFT) in $d$ dimensions whose only light single trace operator is the stress tensor. This contrasts with more standard cases like Type IIB string theory on $AdS_5\times S^5$ dual to $\mathcal{N}=4$ Super-Yang-Mills, where the CFT has light single trace operators for each Kaluza-Klein mode on $S^5$. We compute the 1-loop correction to the pure AdS$_{d+1}$ theory in a small Planck length expansion, which is dual to the large central charge expansion in the CFT. We find that this correction saturates the most general non-perturbative conformal bootstrap bounds on this correlator in the large central charge regime for $d=3,4,6$, while the 1-loop correction to CFTs with string/M-theory duals all lie inside the allowed region.
\end{abstract}

\maketitle
\nopagebreak

\section{Introduction}

The AdS/CFT duality relates quantum gravity on Anti-de Sitter (AdS) space in $d+1$ dimensions times a compact spacetime factor, to certain supersymmetric CFTs in $d$ dimensions \cite{Maldacena:1997re}. In the simplest examples, the compact space is simply a sphere with a similar radius as AdS, and the CFT is maximally supersymmetric. Compactifying the graviton on the sphere generates an infinite tower of Kaluza-Klein (KK) modes in AdS, which are dual to light single trace operators in the CFT. It is an open question if holographic duals exist where the radius of the sphere is parametrically smaller than that of AdS, so that these extra dimensions would be small (See \cite{Alday:2019qrf,Gopakumar:2022kof} for a recent discussion). In the most extreme case, there would simply be no compact factor at all, and the only single trace operators in the dual CFT would be the stress tensor multiplet. No such pure AdS theory has been constructed, despite much effort \cite{Witten:2007kt,Maloney:2007ud,Hellerman:2009bu,Keller:2014xba,Collier:2016cls,Afkhami-Jeddi:2019zci,Hartman:2019pcd,Maxfield:2020ale,Afkhami-Jeddi:2020ezh,Maloney:2020nni}.

We will address this question by studying the stress tensor four-point function, which is dual to scattering of gravitons in the bulk, in maximally supersymmetric CFTs in $d=3,4,6$ dimensions. Consider the large central charge $c$ expansion of this correlator, where $c$ is defined as the coefficient of the stress-tensor two-point function, and is related to the bulk as 
\es{ctoBulk}{
c\sim (L_\text{AdS}/\ell_\text{Planck})^{D-2}\,,
}
where $L_\text{AdS}$ is the radius of the AdS$_{d+1}$ factor, and $\ell_\text{Planck}$ is the Planck length of the full $D$-dimensional bulk spacetime, including a possible compact factor. We can define the correlator $\mathcal{G}$ in any such theory to any order in $1/c$ as
\es{G}{
\cG&=\cG^{(0)}+c^{-1}\cG^R+c^{-2}(\cG^{R|R}+\kappa \cG^{R^4})+\dots\\
&\dots+c^{-\frac{D+4}{D-2}}\cG^{R^4}+c^{-\frac{D+8}{D-2}}\cG^{D^4R^4}+\dots\,,
}
where in the first line we wrote the tree level supergravity term $\mathcal{G}^R$ and the 1-loop term $\cG^{R|R}$ with supergravity vertices $R$, while in the second line we wrote tree level higher derivative corrections that are allowed by supersymmetry \footnote{In CFTs dual to M-theory the lowest correction $R^4$ scales as $c^{-5/3}$, and was computed in \cite{Chester:2018aca,Chester:2018dga}. In CFTs dual to string theory, this coefficient scales like $c^{-7/4}$ at finite string coupling, and was computed for Type IIA in \cite{Binder:2019mpb}, and Type IIB in \cite{Chester:2019jas}. The $D^4R^4$ term has also been computed for M-theory in \cite{Binder:2018yvd}, and for Type IIB in \cite{Chester:2020vyz}.}. The expansion also includes 1-loop terms with such higher derivative vertices, as well as higher loop terms \footnote{The distinction between tree and loop is ambiguous, since $c\sim (L_\text{AdS}/\ell_\text{Planck})^{D-2}$ is the only expansion parameter, but at low orders for some $D$ they can be distinguished by the powers of $1/c$}. The $\mathcal{G}^{R|R}$ term has an $\mathcal{G}^{R^4}$ type contact term with coefficient $\kappa$ as long as the scaling of the $R^4$ tree level term is smaller than $R|R$, which is the case for string and M-theory with $D=10,11$ \footnote{This contact term has been fixed for  M-theory on $AdS_4\times S^7/\mathbb{Z}_k$ \cite{Alday:2021ymb,Alday:2022rly} and Type IIB on $AdS_5\times S^5/\mathbb{Z}_k$  \cite{Chester:2019pvm,Alday:2021vfb}  for $k=1,2$.}, respectively, but is not for the pure AdS$_{d+1}$ theory where $D=d+1$ and $d=3,4,6$. All tree and loop supergravity terms $\mathcal{G}^{R|R|\dots}$ can be computed iteratively using the analytic bootstrap \cite{Rastelli:2017udc,Aharony:2016dwx}, but to fix the higher derivative corrections as well as loop contact terms such as $\kappa \cG^{R^4}$, we need a UV completion like string/M-theory. These terms only affect CFT data with finite spin \cite{Heemskerk:2009pn}, so at any given order in $1/c$ we can unambiguously determine an infinite set of CFT data for AdS$_{d+1}$ duals with any (or no) compact factor. Whether or not a pure AdS$_{d+1}$ theory is also defined non-perturbatively in $c$ is a separate question that we will address in the conclusion.

The tree level supergravity correction $\mathcal{G}^{R}$ at order $1/c$ is unaffected by a compact spacetime factor \cite{Rastelli:2017udc,Rastelli:2017ymc,Zhou:2017zaw,Alday:2020dtb}, but higher loop terms starting with $\mathcal{G}^{R|R}$ at order $1/c^2$ are sensitive to the number of KK modes \cite{Aharony:2016dwx}. We will compute this 1-loop term for pure AdS$_{d+1}$ theories in $d=3,4,6$ using the analytic bootstrap, which allows us to extract all CFT data to $O(c^{-2})$. We then can compare this $O(1/c^2)$ data to non-perturbative numerical bootstrap bounds \cite{Beem:2013qxa,Beem:2015aoa,Chester:2014fya,Chester:2014mea}, which apply to any maximally supersymmetric CFT, and can be computed for any $c$. We find that for all $d=3,4,6$, the pure AdS$_{d+1}$ 1-loop correction precisely saturates the bootstrap bounds in the large $c$ regime.

The 1-loop correction has also been computed for maximally supersymmetric CFTs with string/M-theory duals. In 3d, these CFTs are $U(N)_{k}\times U(N)_{-k}$ ABJM theory with $k=1,2$, which is dual to M-theory on $AdS_4\times S^7/\mathbb{Z}_k$ with $c\sim N^{3/2}$ \footnote{The $U(N)_{2}\times U(N+1)_{-2}$ theory also has maximal supersymmetry, but this shift of the gauge factor does not matter in the large $N$ limit. When $k>2$, the theory has $\mathcal{N}=6$ supersymmetry.} \cite{Aharony:2008ug}. In 4d, they are $\mathcal{N}=4$ super-Yang-Mills (SYM) with gauge group $SU(N)$ or $SO(N)$ \footnote{The $USp(2N)$ gauge group is also allowed, but is similar to $SO(N)$ in the large $N$ limit.}, which is dual to Type IIB string theory on $AdS_5\times S^5$ or $AdS_5\times S^5/\mathbb{Z}_2$ with $c\sim N^2$ \cite{Maldacena:1997re}, respectively. In 6d, they are $A_{N-1}$ or $D_N$ $(2,0)$ theories \cite{Witten:1995zh} \footnote{There are also $(2,0)$ theories constructed from exceptional groups, but these do not have a large $N$ limit.}, which are dual to $AdS_7\times S^4$ or $AdS_7\times S^4/\mathbb{Z}_2$ with $c\sim N^3$ \cite{Witten:1998xy,Aharony:1998rm}, respectively. The 1-loop corrections were computed in these various cases in \cite{Alday:2017xua,Aprile:2017bgs,Alday:2020tgi,Alday:2021ymb,Alday:2021vfb,Alday:2022rly}. In all cases, we find that these corrections lie inside the allowed region of the bootstrap bounds for the same regime of large $c$ where the pure AdS$_{d+1}$ theory saturates the bound.

The rest of this paper is organized as follows.  In Section \ref{sec:stress}, we review the constraints of maximal superconformal symmetry on the stress tensor four-point function for $d=3,4,6$. In Section \ref{sec:1loop} we consider the large $c$ expansion of this correlator and compute the 1-loop correction to pure AdS$_{d+1}$ supergravity. In Section \ref{sec:bootstrap} we compare this correction, and the previously computed 1-loop corrections for string/M-theory duals, to non-perturbative numerical conformal bootstrap bounds in the large $c$ regime. We end with a discussion of our results in Section~\ref{sec:discussion}.

\section{Stress tensor correlator}\label{sec:stress}

We begin by reviewing the constraints of maximal supersymmetry in $d=3,4,6$ on the stress tensor correlator. We consider the superconformal primary $S( x)$, which is a scalar with $\Delta=d-2$ that transforms in the symmetric traceless representation of the R-symmetry group $SO(8)_R$, $SO(6)_R$, and $SO(5)_R$ for 3d, 4d, and 6d, respectively. Conformal and R-symmetry fixes the four-point function to take the form
\es{4point}{
&\langle S( x_1,Y_1)S( x_2,Y_2)S( x_3,Y_3)S( x_4,Y_4) \rangle=\\
&\qquad\qquad\qquad\frac{(Y_1\cdot Y_2)^2(Y_3\cdot Y_4)^2}{| x_{12}|^{2(d-2)}| x_{34}|^{2(d-2)}}\mathcal{G}(U,V;\sigma,\tau)\,,
}
where we define the cross ratios
 \es{uvsigmatauDefs}{
&  U \equiv \frac{{x}_{12}^2 {x}_{34}^2}{{x}_{13}^2 {x}_{24}^2} \,, \qquad
   V \equiv \frac{{x}_{14}^2 {x}_{23}^2}{{x}_{13}^2 {x}_{24}^2}  \,, \\
&   \sigma\equiv\frac{(Y_1\cdot Y_3)(Y_2\cdot Y_4)}{(Y_1\cdot Y_2)(Y_3\cdot Y_4)}\,,\qquad \tau\equiv\frac{(Y_1\cdot Y_4)(Y_2\cdot Y_3)}{(Y_1\cdot Y_2)(Y_3\cdot Y_4)} \,,
 }
 with $x_{ij}\equiv x_i-x_j$, and $Y_i$ are null polarization vectors that encode the R-symmetry indices. The constraints from supersymmetry are given by the superconformal Ward identities \cite{Dolan:2004mu}, which can be satisfied by expanding $ \mathcal{G}$ in superconformal blocks as \footnote{In 4d and 6d, we can also satisfy these Ward identities by writing $ \mathcal{G}(U,V;\sigma,\tau)$ in terms of a differential operator $\Upsilon(U,V,\partial_U,\partial_V,\sigma,\tau)$ acting on a reduced correlator $\mathcal{H}(U,V)$, which is then an R-symmetry singlet. }
\es{SBDecomp}{
     \mathcal{G}(U,V;\sigma,\tau)=\sum_{\mathcal{M}}\lambda^2_\mathcal{M}\mathfrak{G}_\mathcal{M}(U,V;\sigma,\tau)\,,
}
where ${\cal M}$ runs over all the supermultiplets appearing in the $S \times S$ OPE, the $\lambda^2_{\mathcal{M}}$ are the squared OPE coefficients for each such supermultiplet $\mathcal{M}$, and the explicit form of the superblocks can be found for each $d$ in \cite{Dolan:2004mu,Beem:2016wfs,Beem:2015aoa,Chester:2014fya}. In Appendix \ref{sec:multiplets}, for each $d$ we summarize the multiplets $\mathcal{M}$ that appear, which we label by the scaling dimension $\Delta$, the spin $\ell$, and the R-symmetry representation of the superprimary. We exclude free theory multiplets, which for $d=4,6$ restricts us to interacting theories \footnote{In 3d, the free theory multiplet is identical to the unitarity bound of the long multiplet, so cannot be excluded kinematically}. The $S\times S$ OPE includes long multiplets in the singlet of the R-symmetry group with even spin $\ell$ and scaling dimension $\Delta> d-2+\ell$, as well as protected multiplets such as the stress tensor with fixed $\Delta$. The stress tensor $\lambda^2$ is fixed by the conformal Ward identity \cite{Osborn:1993cr} to be inversely proportional to the central charge coefficient $c$ of the stress tensor two-point function:
\es{c}{
  \lambda^2_\text{stress}\propto1/c\,,
}
where the proportionality constant is fixed in 4d so that $c$ is the conformal anomaly \cite{Beem:2016wfs}, in 6d so that a free tensor multiplet has $c=1$ \cite{Beem:2015aoa}, and in 3d so that the free theory has $c=16$ \cite{Chester:2014fya}. In 4d and 6d, the existence of a protected 2d chiral algebra \cite{Beem:2013sza} fixes $\lambda^2_{\mathcal{M}}\propto1/c$ for certain protected multiplets, while the remaining protected multiplets $\mathcal{M}_\text{prot}$ have $\lambda^2$ that remain unconstrained. 

An important non-perturbative constraint on the four-point function can be derived by swapping $1\leftrightarrow3$ in \eqref{4point}, which yields the crossing equations 
\es{crossing}{
\cG(U,V;\sigma,\tau)=\frac{U^{d-2}}{V^{d-2}}\tau^2 \cG(V,U;\sigma/\tau,1/\tau)\,,
}
which we will now use to constrain the correlator.

\section{One-loop from tree level}\label{sec:1loop}

We will now restrict to the pure AdS$_{d+1}$ theory, and consider the large $c$ expansion of the correlator $\cG$ shown in \eqref{G}, where we expand long multiplet CFT data as
\es{Hlarge}{
\Delta_{n,\ell}&=2(d-2)+2n+\ell+\gamma^{R}_{n,\ell}/c+\gamma^{R|R}_{n,\ell}c^2+\dots\,,\\
\lambda_{n,\ell}^2&=(\lambda^{(0)}_{n,\ell})^2+(\lambda^{R}_{n,\ell})^2/c+(\lambda^{R|R}_{n,\ell})^2/c^2+\dots\,.\\
}
A similar expansion exists for the OPE coefficients of the protected operators, although of course their scaling dimensions are fixed. The long multiplets that appear in \eqref{Hlarge} are all double trace operators $[SS]_{n,\ell}$ of the schematic form
\es{longs}{
[SS]_{n,\ell}=S\Box^n\partial_{\mu_1}\dots\partial_{\mu_\ell}S\,,
}
with $\Delta^{(0)}_{n,\ell}=2(d-2)+2n+\ell$ in the $c\to\infty$ generalized free field theory (GFFT). Note that if the bulk theory had a compact factor, e.g. $AdS_5\times S^5$, then we could use the higher KK modes to construct more such long operators, which would be degenerate in the GFFT and thus mix in the $1/c$ expansion. The GFFT and tree correlators, which are insensitive to the bulk factor, were computed in each $d$ in \cite{Dolan:2001tt,Zhou:2017zaw,Heslop:2004du,Arutyunov:2002ff} and used to extract tree level data, which we summarize in Table \ref{summary}. For theories with higher KK modes, we can only extract the average long multiplet anomalous dimensions $\langle \lambda^2_{n,\ell} \gamma_{n,\ell}^R\rangle$, due to the degeneracy at GFFT. For protected multiplets, we can obtain the unique CFT data for all such large $c$ theories.

At 1-loop level, we can expand the superblock expansion \eqref{SBDecomp} to get
\es{1loopG}{
&\mathcal{G}^{R|R}=\sum_{n=0}^\infty\sum_{\ell\in\text{Even}}\Big[\frac18{(\lambda^{(0)}_{n,\ell})^2(\gamma^{R}_{n,\ell})^2}\log^2U\\
&\hspace{-.1in}+{(\lambda^{(0)}_{n,\ell})^2\gamma^{R|R}_{n,\ell}}\frac{\log U}{2}+\dots\Big]\mathfrak{G}_{n,\ell}+\hspace{-.05in}\sum_{\cM_\text{prot}}(\lambda^{R|R}_\cM)^2\mathfrak{G}_\cM\,,
}
where the ellipses refers to other other combinations of tree and loop data, and recall that $\cM_\text{prot}$ denotes protected multiplets whose OPE coefficients are not $1/c$ exact. The significance of the $\log^2U$ term is that it is the only term at this order that has a double discontinuity (DD) as $U\to0$  \footnote{This is true for every known maximally supersymmetric CFT in $d=3,4,6$ except the $U(N)_1\times U(N)_{-1}$ ABJM theory, for which additional contributions come from odd twist long multiplet OPE coefficients, which can also be computed from tree level data. See \cite{Alday:2022rly} for more details.}. The Lorentzian inversion formula \cite{Caron-Huot:2017vep} shows that all CFT data with sufficiently large $\ell$ can be extracted from the DD as $V\to0$, so we can obtain this DD from the $\log^2U$ terms after applying crossing \eqref{crossing}. For instance, we can compute the 1-loop correction to the OPE coefficient of 3d protected multiplets ${(A,+)_\ell}$ as 
\es{ApInversion}{
&(\lambda^{R|R}_{(A,+)_\ell})^2 =\frac{12 (2 \ell+5) \Gamma (\ell+3)^4}{\Gamma \left(\ell+\frac{5}{2}\right)^2
   \Gamma \left(\ell+\frac{7}{2}\right)^2}\\
   &\times \int_0^1 \frac{d \bar z}{\bar z}    g_{\ell+4,\ell+2}(\bar z) \text{dDisc}[ {\cal G}^{[0040]}(z\bar z,1-\bar z,)\vert_z ] \,,
}
where ${\cal G}^{[0040]}|_z$ is the leading twist term in the highest weight representation of $SO(8)_R$, we define the lightcone blocks $g_{\Delta,\ell}(z)$ in Appendix \ref{sec:inverse}, and we introduce the variables $U=z\bar z$ and $V=(1-z)(1-\bar z)$. We compute $ \text{dDisc}$ acting on $\log^2 V\sim \log^2(1-\bar z)$ as
\es{ddisc}{
 {\rm dDisc}\,[ f(z,\bar z)\log^2{1-\bar z} ] = 4\pi^2 f(z,\bar z)\,,
}
where we assume $f(z,\bar z)$ is analytic as $\bar z\to1$ (i.e. $V\to0$ in a small $U$ expansion). We give the inversion formulae for the other CFT data in Appendix \ref{sec:inverse}. Note that in the string/M-theory cases, the inversion formula does not converge for low spins, which corresponds to the existence of the contact terms $\kappa \mathcal{G}^{R^4}$ in \eqref{G}. In the pure AdS$_{d+1}$ case we do not have such contact terms as discussed above, so we can in fact extract all CFT data at 1-loop order. 

To apply these inversion formulae, we need to compute the $\log^2U$ terms in \eqref{1loopG} for finite $U$, expand to leading in $U$ in the crossed channel \eqref{crossing}, and perform the integral of the resulting resummed $V\sim 1-\bar z$ expression. We compute the $\log^2U$ terms in a small $U$ expansion using the ansatz
\es{DD}{
&\frac18\sum_{n=0}^\infty\sum_{\ell\in\text{Even}}{(\lambda^{(0)}_{n,\ell})^2(\gamma^{R}_{n,\ell})^2}\mathfrak{G}_{n,\ell}=\sum_{n=0}^\infty U^{d-2+n} \Big[p_1\\
&+p_2\log V+p_3\log^2V+p_4 \text{Li}_2(1-V)\Big]+\dots\,,
}
where here we showed the singlet channel, while the dots denote the other R-symmetry channels $R$ that will start at higher powers of $U$ and have nontrivial $\sigma,\tau$ dependence given by the structures $Y_R(\sigma,\tau)$, as given in Eq. B.14 of \cite{Nirschl:2004pa}. The coefficients $p_i$ are polynomials in $V$ divided by monomials in $V$. We then perform crossing, expand to leading order $U$, and resum the expansion in $V\sim1-\bar z$ to get the relevant DDs. The final expressions are inverse trigonometric functions of $\bar z$ times high degree polynomials in $\bar z$, whose explicit form we give in the attached \texttt{Mathematica} file. We then plug these into the inversion formula to obtain the 1-loop correction to CFT data. For the lowest spin in each multiplet we find
\es{3dresults}{
(\lambda^{R|R}_{(B,+)})^2&=793.76\,,\qquad (\lambda^{R|R}_{(A,+)_0})^2=97.766\,,\\
(\lambda^{R|R}_{(B,2)})^2&=3968.8\,,\qquad\; (\lambda^{R|R}_{(A,2)_1})^2=570.50\,,\\
\gamma^{R|R}_{0,0}&=14043\,,\qquad\qquad\, \Delta^{R|R}_{3d,2}=2713.6\,,\\
}
where here we show 5 digits of precision, but we can compute arbitrary precision. In 4d, the only nontrivial data is the anomalous dimensions, which were already computed for pure AdS$_5$ in \cite{Alday:2017xua} for $\ell\geq0$:
\es{4dresults}{
\gamma^{R|R}_{0,\ell}=\frac{24 \left(7 \ell^5+116 \ell^4+725 \ell^3+2044 \ell^2+2292 \ell+288\right)}{(\ell+1)^2 (\ell+2) (\ell+3) (\ell+4) (\ell+5) (\ell+6)^3}\,.
}
In 6d, we compute the lowest few spins for the multiplets with non-trivial $1/c$ expansions to get
\es{6dresults}{
(\lambda^{R|R}_{\mathcal{B}[0,2]_1})^2&=-4.2372\,,\qquad (\lambda^{R|R}_{\mathcal{B}[0,2]_3})^2=-0.1531\,,\\
\gamma^{R|R}_{0,0}&=-54695\,,\qquad\qquad\; \Delta^{R|R}_{6d,2}=-258.79\,,\\
\gamma^{R|R}_{0,4}&=-17.157\,,\qquad\; \,(\lambda^{R|R}_{\mathcal{D}[0,4]})^2=-822.70\,.
}

\section{Numerical conformal bootstrap}\label{sec:bootstrap}

We will now compare these 1-loop corrections to the numerical bootstrap bounds on CFT data in the stress tensor correlator for $d=3,4,6$, which were computed for $d=3,4$ in \cite{Alday:2021ymb,Alday:2021vfb}, and which we compute now for 6d following \cite{Beem:2015aoa}. These bounds come from optimizing the infinite set of constraints imposed by the crossing equations \eqref{crossing} on the superblock expansion in \eqref{SBDecomp}, for more details in each case see the original works \cite{Beem:2013qxa,Beem:2015aoa,Chester:2014fya}, and \cite{Poland:2018epd,Chester:2019wfx,Simmons-Duffin:2016gjk,Poland:2022qrs} for recent reviews. The convergence of these bounds is monotonic and given by the parameter $\Lambda$ originally defined in \cite{Chester:2014fya}, which counts how many derivatives are used in the expansion of conformal blocks around the crossing symmetric point \footnote{For comparison, the most precise Ising model bounds were computed with $\Lambda=43$ in \cite{Landry:2019qug}, while all the bounds shown here use at least twice that precision.}. These bounds apply to any theory with maximal supersymmetry in the given $d$ and are computed as a function of $c$, which is related to the stress tensor OPE coefficient as in \eqref{c}. Since these bounds are non-perturbative in $c$, we will look at the large $c$ regime where we expect the $1/c$ expansion of the previous section to be good. The large $c$ expansion of CFT data is asymptotic, which means that after a few orders the expansion will actually get worse, unless we look at very large values of $c$. We observe that the $1/c^2$ corrections get smaller relative to $1/c$ tree corrections as the spin increases, which implies that the asymptotic expansion is getting more accurate at this order. We do not want to look at very high spin data, however, because then the difference between each order will be hard to observe. As a compromise, we will focus on the lowest spin CFT data for which the Lorentzian inversion converges for the string/M-theory CFTs. We summarize the comparison of the analytic $1/c$ expansion to fits in the large $c$ regime of the bootstrap bounds in Table \ref{summary} \footnote{A rough diagnostic for the error of these fits is given by how close the $1/c$ tree level correction matches the known answer. The range of $c$ used for the fits was motivated to give such a tree level match, such that the 1-loop term is then a prediction.}.

\begin{table}
\centering
\begin{tabular}{c|l|c}
3d:&$\Delta_{0,2}$: Exact&  $4 - 49.931/c + 2713.6/c^2$         \\ 
 & \qquad\; Fit & $3.99996 - 49.82/c + 2619.4 /c^2$\\
\hline
&$\lambda^2_{(A,2)_1}$: Exact&  $9.7523 - 98.764 /c + 570.43 /c^2$         \\ 
 & \qquad\quad\; Fit & $9.7523 - 98.772 /c + 580.443 /c^2$\\
\hline
&$\lambda^2_{(A,+)_0}$: Exact&  $7.1111 + 48.448 /c + 97.768 /c^2$         \\ 
 &  \qquad\quad\;\; Fit & $7.1111 + 48.445 /c + 103.35 /c^2$\\
\hline
4d:&$\Delta_{0,2}$: Exact&  $6 - 1/c + 0.12976/c^2$         \\ 
 & \qquad\; Fit & $6.0000 - 0.99929 /c + 0.14718 /c^2$\\
\hline
6d:&$\Delta_{0,2}$: Exact&  $10 - 10.909 /c - 258.79 /c^2$         \\ 
 & \qquad\; Fit & $10.000 - 11.209 /c + 270.96 /c^2$\\
\hline
&$\Delta_{0,4}$: Exact&  $12 - 3.1648 /c - 17.157 /c^2$         \\ 
 & \qquad\; Fit & $12.000 - 3.1956 /c - 17.832 /c^2$\\
\hline
&$\lambda^2_{\mathcal{B}[02]_1}$: Exact&  $0.75757 - 0.98484 /c - 4.2372 /c^2$         \\ 
 & \qquad\quad\; Fit & $0.75757 - 0.98009 /c - 3.9446 /c^2$\\
\hline
&$\lambda^2_{\mathcal{B}[02]_3}$: Exact&  $0.43076 - 0.15440 /c - 0.15313 /c^2$         \\ 
 &  \qquad\quad\;\; Fit & $0.43076 - 0.15432 /c - 0.17448 /c^2$\\
\hline
\end{tabular}
\caption{Fits of the numerical bootstrap bounds at large $c$, compared to exact $O(1/c^2)$ values for the pure AdS$_{d+1}$ theory for $d=3,4,6$.}
\label{summary}
\end{table} 

\begin{figure*}
	\centering
	\includegraphics[width=\columnwidth]{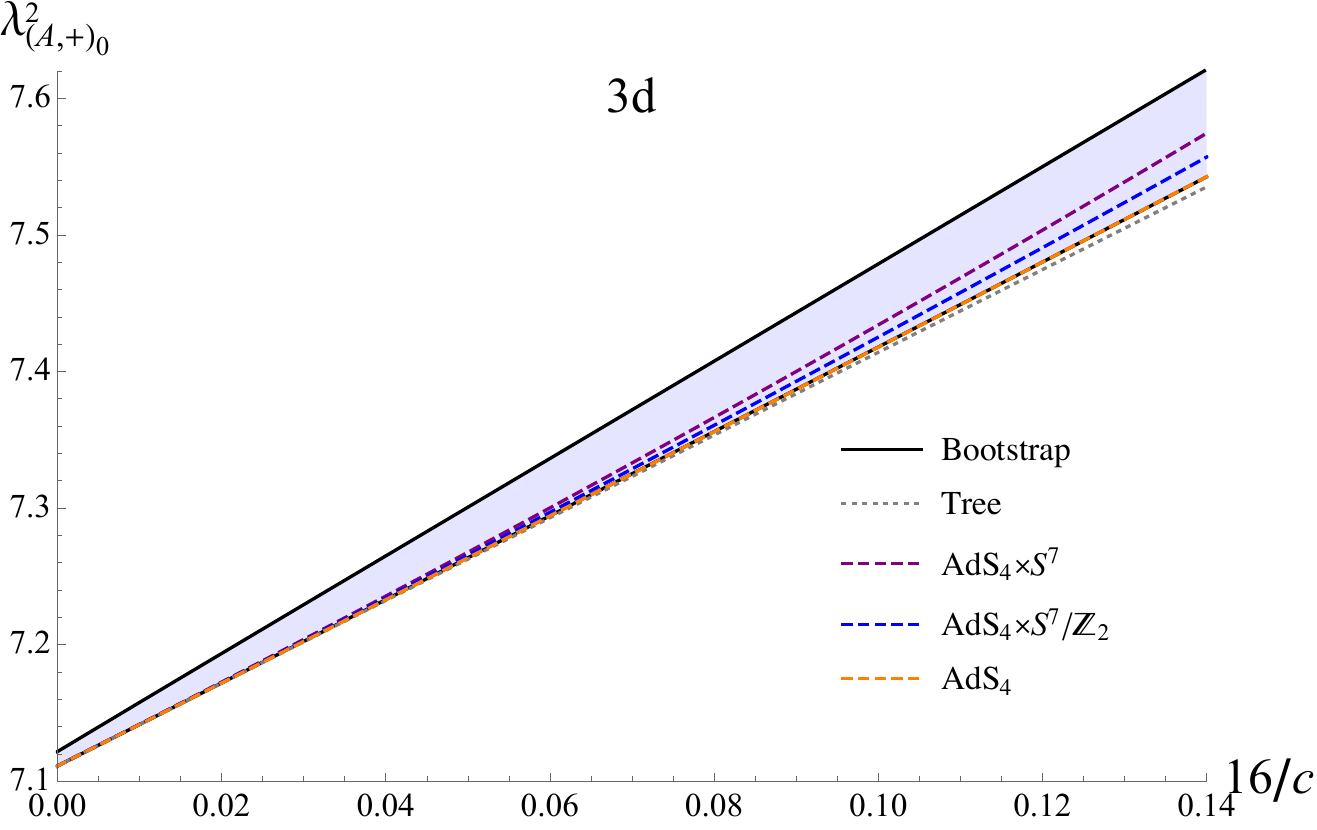}
	\includegraphics[width=\columnwidth]{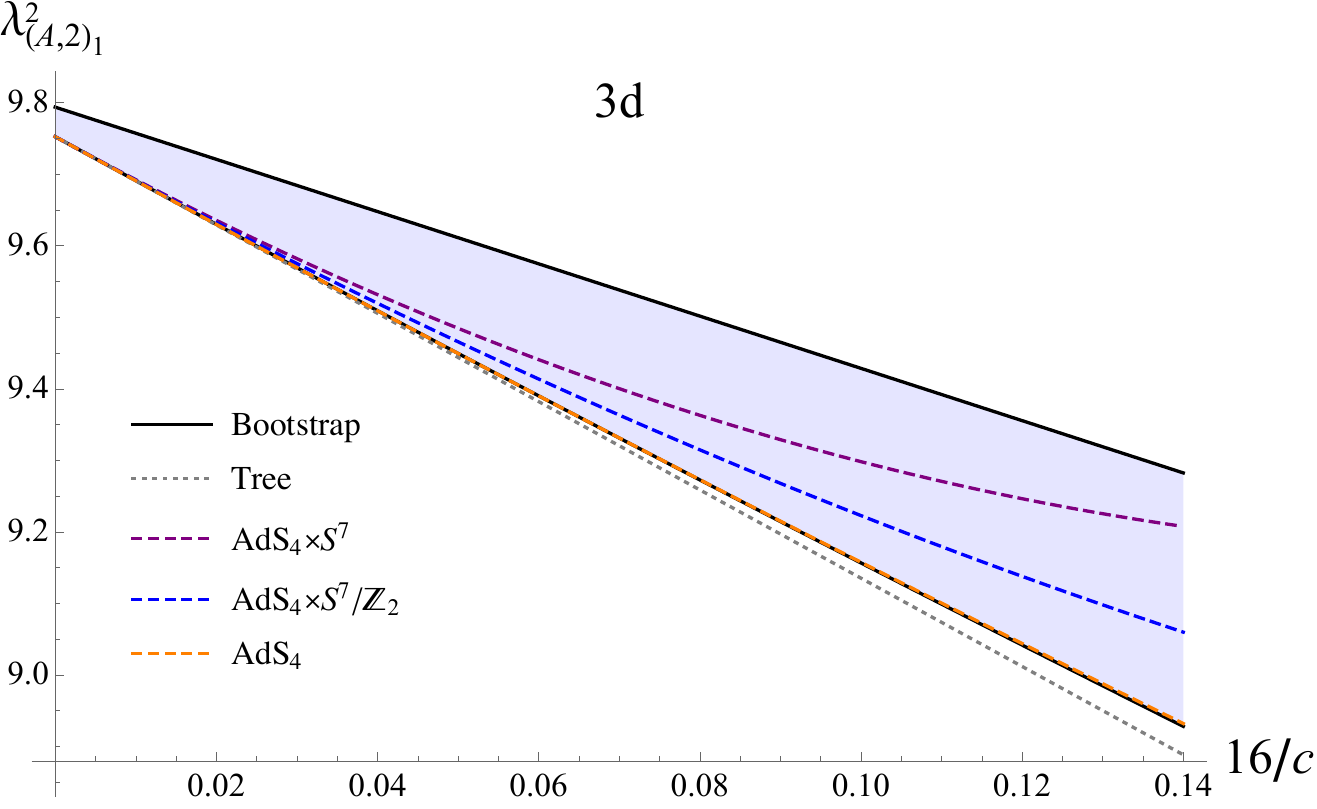}
	\includegraphics[width=\columnwidth]{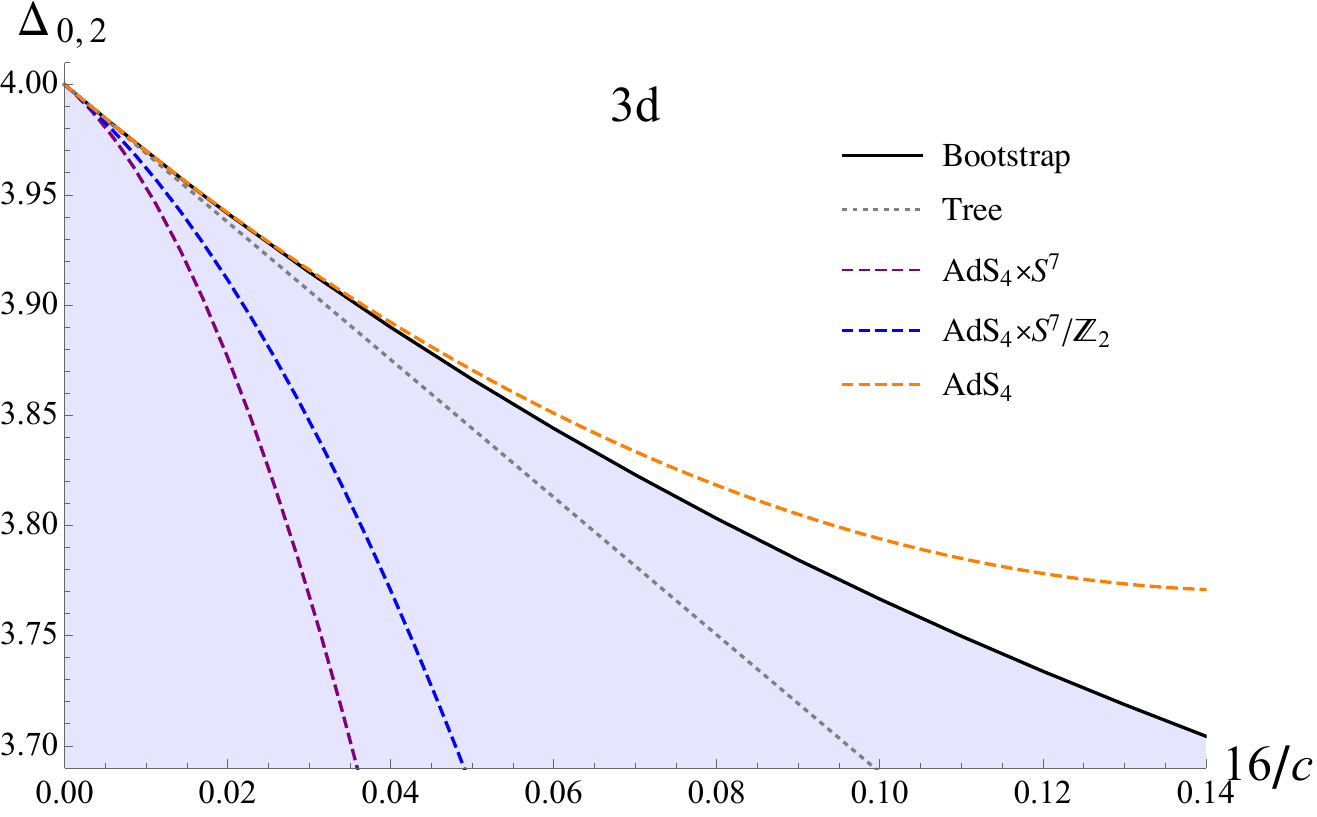}
	\caption{Upper and lower numerical bootstrap bounds (in black) on the $\lambda_{(A,+)_0}^2$ and $\lambda_{(A,2)_1}^2$ OPE coefficients, as well as upper bounds on the scaling dimension $\Delta_{0,2}$ of the lowest dimension spin 2 long multiplet, made with precision $\Lambda=83$. These bounds apply to any 3d $\mathcal{N}=8$ CFT, and are plotted in terms of the stress-tensor coefficient $c$ in the large $c$ regime, where $c=16$ for the free theory. The gray dotted line denotes the large $c$ expansion to order tree level supergravity $O(c^{-1})$, which does not depend on the compact factor in the bulk. The purple, blue, and orange dashed lines also include the 1-loop supergravity correction $O(c^{-2})$ on $AdS_4\times S^7$, $AdS_4\times S^7/\mathbb{Z}_2$, and AdS$_4$, respectively. }
	\label{3dbound}
\end{figure*}

We start with the the bounds on 3d $\mathcal{N}=8$ CFTs, which were computed with $\Lambda=83$. In Figure \ref{3dbound} we show upper and lower bounds on OPE coefficients for the protected $(A,+)_\ell$ and $(A,2)_\ell$ multiplets for the lowest spins $\ell=0$ and $\ell=1$, respectively.  Both upper and lower bounds exist for the OPE coefficients, because their protected scaling dimensions $\Delta=\ell+2$ are separated from the continuum of long multiplets. The lower bounds are the nontrivial bounds in this case, as the upper bounds simply interpolate between the GFFT values at $c\to\infty$ and the free theory values at $c=16$. We also show upper bounds on the lowest dimension scaling dimension $\Delta_{0,\ell}$ of the long multiplet with spin $\ell=2$. We compare these bounds to the 1-loop data for the pure AdS$_4$ theory as given in \eqref{3dresults}, as well as for the $AdS_4\times S^7$ and $AdS_4\times S^7/\mathbb{Z}_2$ theories \cite{Alday:2021ymb,Alday:2022rly}, which we review in Appendix \ref{sec:tree}. We find that the pure AdS$_4$ 1-loop correction at $1/c^2$ noticeably improves the universal tree correction at $1/c$ and approximately saturates the numerical bounds, unlike the $AdS_4\times S^7$ and $AdS_4\times S^7/\mathbb{Z}_2$ 1-loop corrections, which lie inside the allowed region. 

\begin{figure}
	\centering
	\includegraphics[width=\columnwidth]{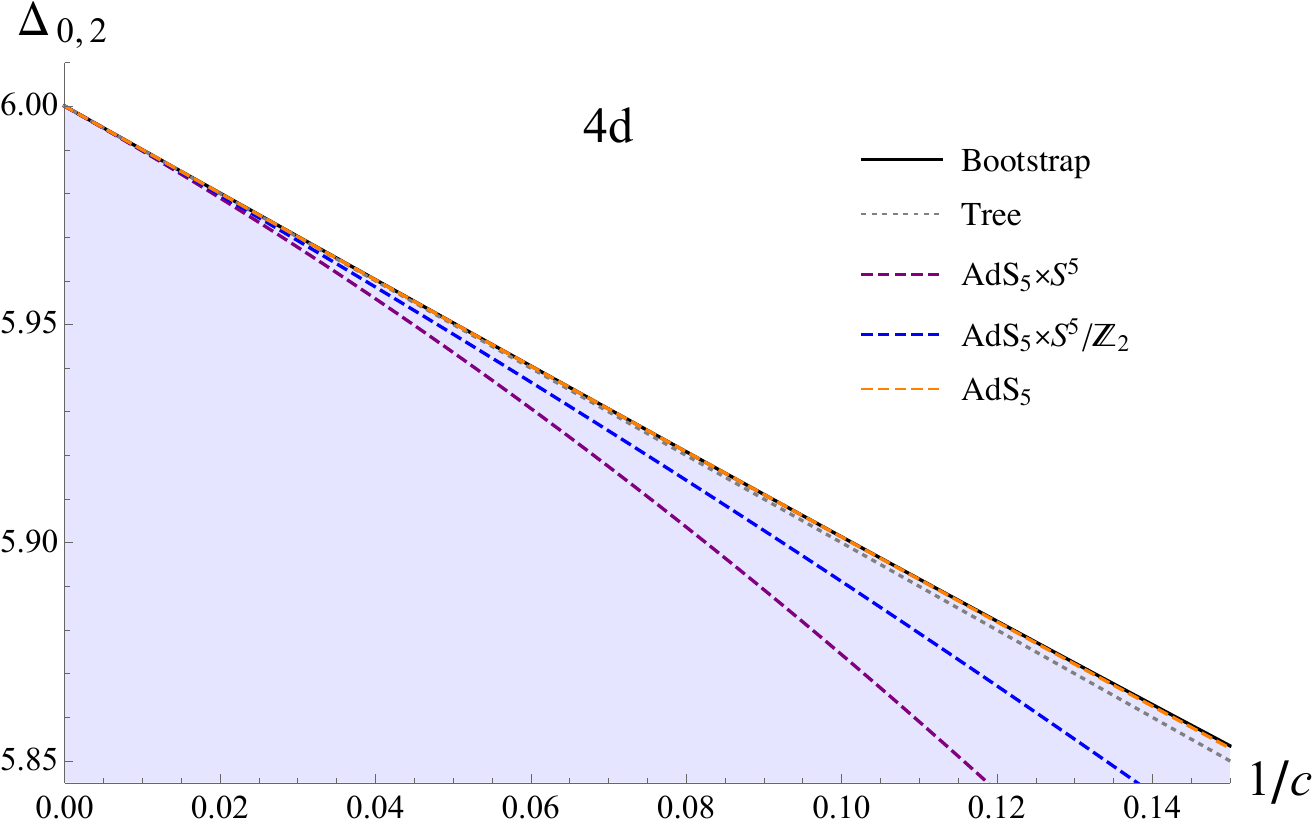}
	\caption{Upper bounds (in black) on the scaling dimension $\Delta_{0,2}$ of the lowest dimension spin 2 long multiplet, made with precision $\Lambda=123$. These bounds apply to any interacting 4d $\mathcal{N}=4$ CFT, and are plotted in terms of the stress-tensor coefficient $c$ in the large $c$ regime, where $c=3/4$ for the minimal interacting theory $SU(2)$ SYM. The gray dotted line denotes the large $c$ expansion to order tree level supergravity $O(c^{-1})$, which does not depend on the compact factor in the bulk. The purple, blue, and orange dashed lines also include the 1-loop supergravity correction $O(c^{-2})$ on $AdS_5\times S^5$, $AdS_5\times S^5/\mathbb{Z}_2$, and AdS$_5$, respectively.}
	\label{4dbound}
\end{figure}

Next, we consider the bounds on 4d $\mathcal{N}=4$ CFTs, which were computed with $\Lambda=123$. In Figure \ref{4dbound} we show upper bounds on the lowest dimension scaling dimension $\Delta_{0,\ell}$ of the long multiplet with spin $\ell=2$. We compare these bounds to the 1-loop data for the pure AdS$_5$ theory as given in \eqref{4dresults}, as well as for the $AdS_5\times S^5$ and $AdS_5\times S^5/\mathbb{Z}_2$ theories \cite{Aprile:2017bgs,Alday:2017xua,Alday:2021vfb}, which we review in Appendix \ref{sec:tree}. Again, we find that the pure AdS$_5$ 1-loop correction noticeably saturates the numerical bounds relative to the tree, $AdS_5\times S^5$, or $AdS_5\times S^5/\mathbb{Z}_2$ expressions. The correction is particularly striking in this case, as the tree level correction lies below the upper bound, and only the pure AdS$_5$ 1-loop correction is positive. 

\begin{figure*}
	\centering
	\includegraphics[width=\columnwidth]{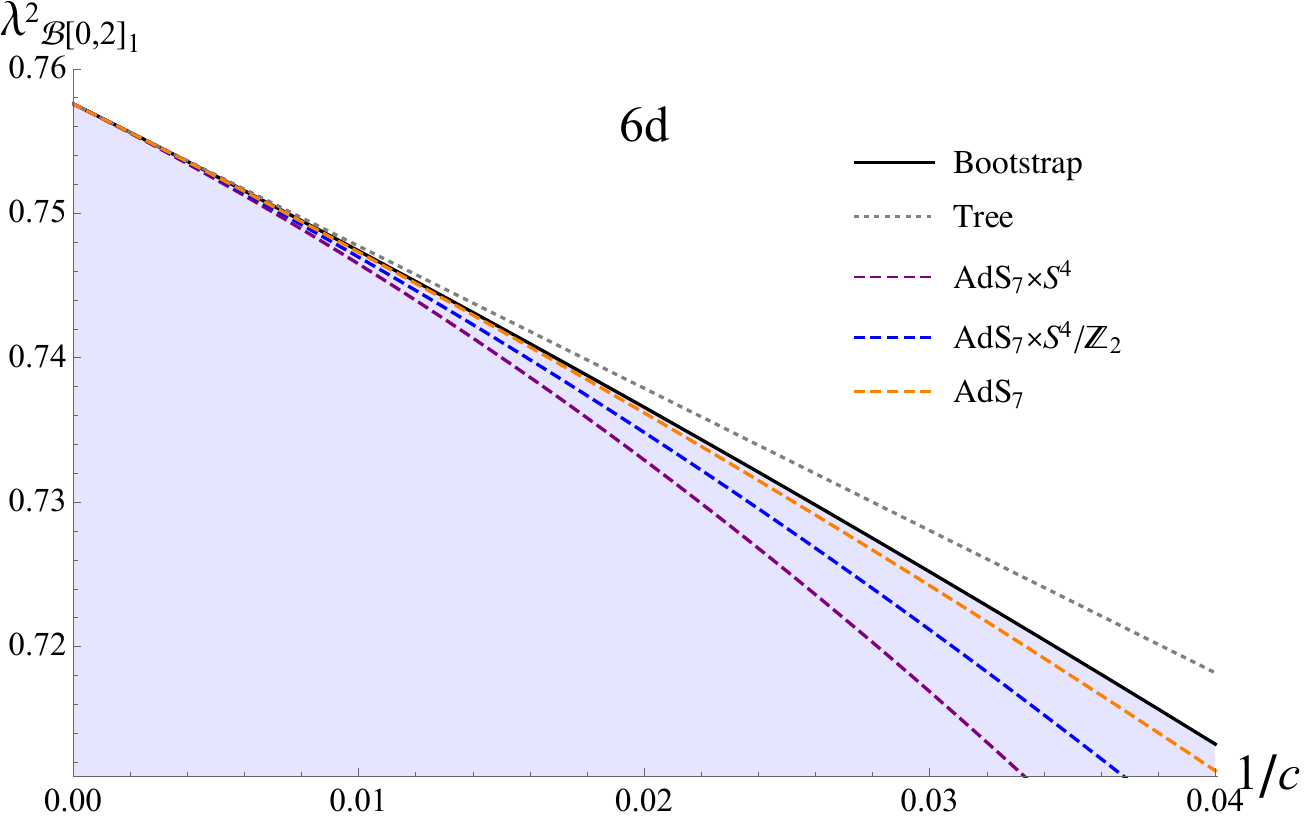}
	\includegraphics[width=\columnwidth]{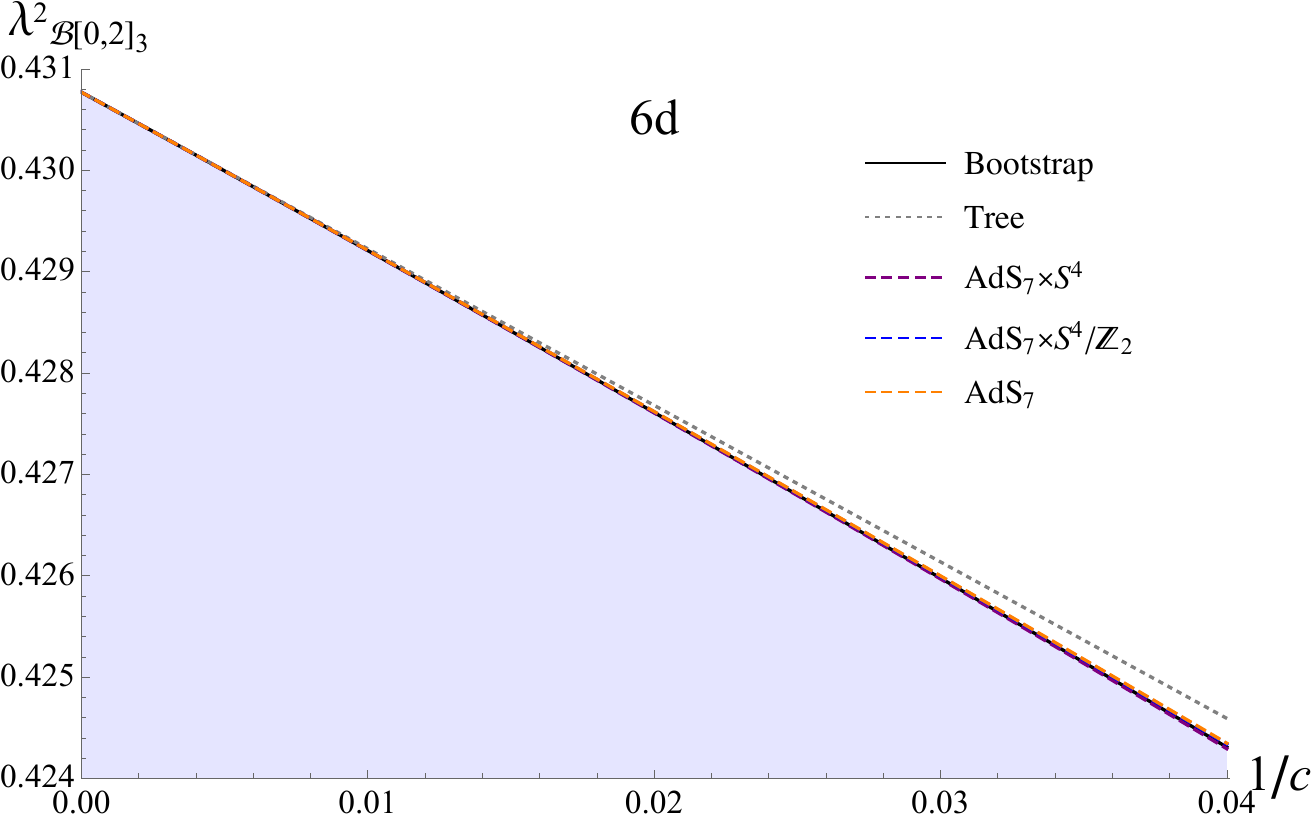}
	\includegraphics[width=\columnwidth]{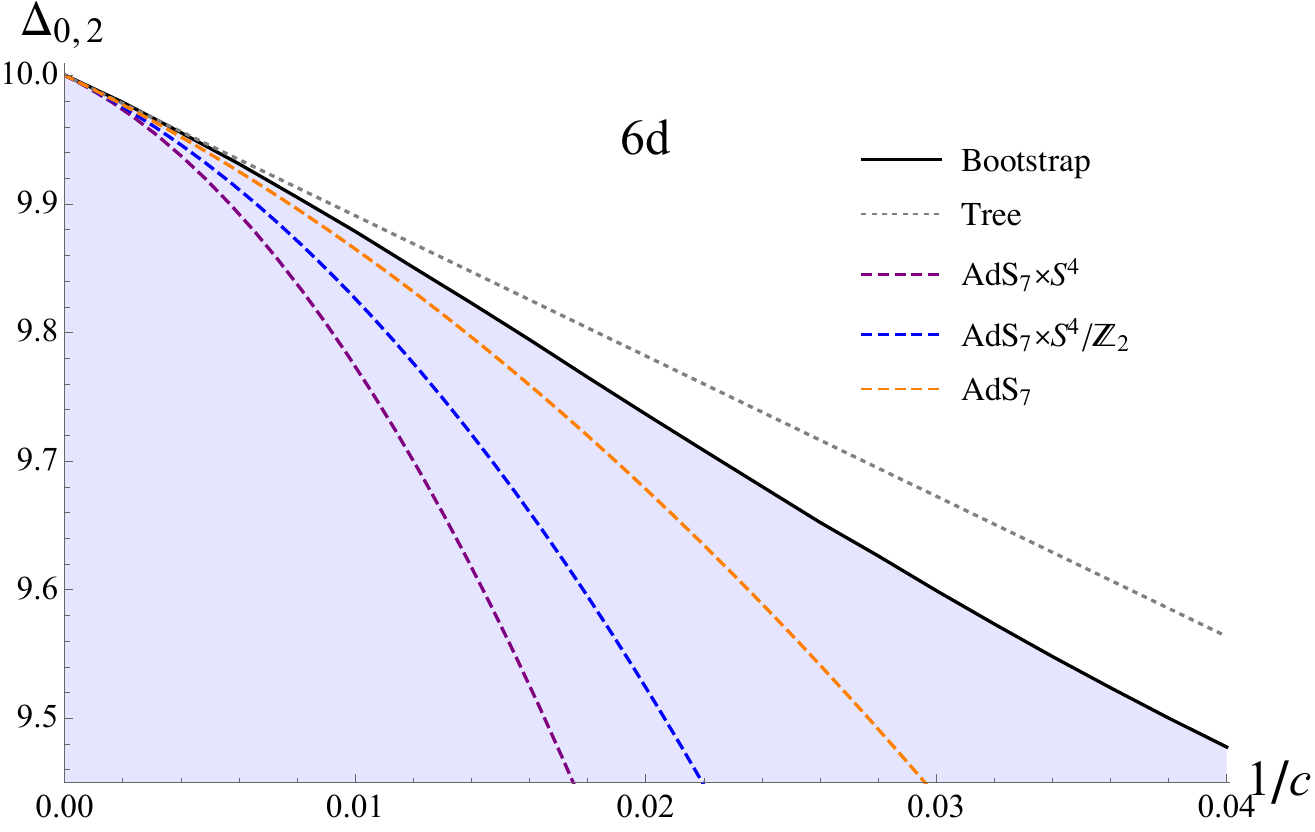}
	\includegraphics[width=\columnwidth]{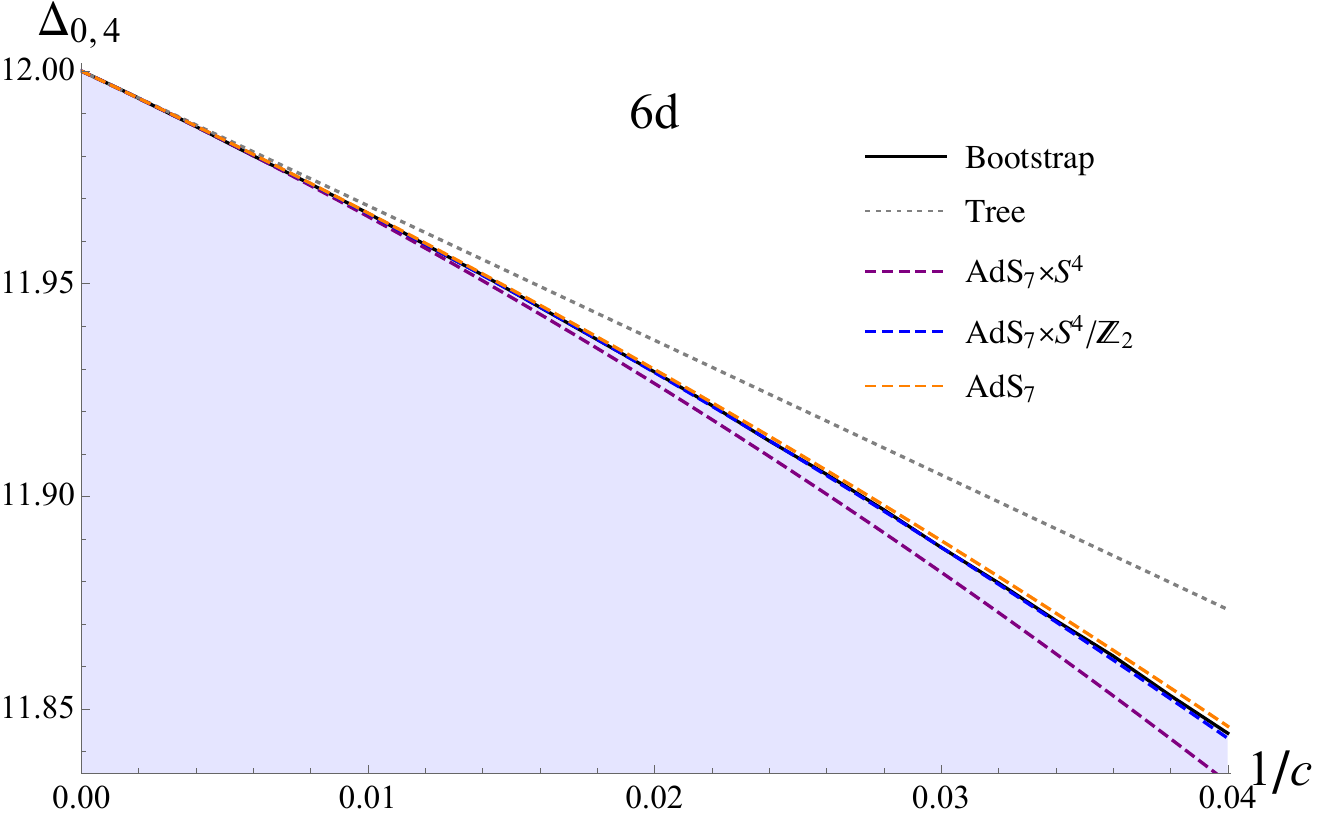}
	\caption{Upper and lower numerical bootstrap bounds (in black) on the $\lambda_{\mathcal{B}[02]_\ell}^2$ OPE coefficients  for $\ell=1,3$, as well as upper bounds on the scaling dimension $\Delta_{0,\ell}$ of the lowest dimension $\ell=2,4$ long multiplets, made with precision $\Lambda=91$. These bounds apply to any interacting 6d $(2,0)$ CFT, and are plotted in terms of the stress-tensor coefficient $c$ for $c\geq25$, which is the value for the minimal interacting theory $A_1$. The gray dotted line denotes the large $c$ expansion to order tree level supergravity $O(c^{-1})$, which does not depend on the compact factor in the bulk. The purple, blue, and orange dashed lines also include the 1-loop supergravity correction $O(c^{-2})$ on $AdS_7\times S^4$, $AdS_7\times S^4/\mathbb{Z}_2$, and AdS$_7$, respectively. }
	\label{6dbound}
\end{figure*}

Finally, we consider the bounds on 6d $(2,0)$ CFTs, which we computed with $\Lambda=91$ \footnote{See \cite{Lemos:2021azv} for a recent numerical bootstrap study of this correlator that compared the bounds to the finite $c$ inversion formula iteratively acted on the protected CFT data.}. The bootstrap is generically less converged as $d$ increases, so in this case we show bounds on two low values of spin for each nontrivial multiplet to show the improvement of the match. Since $c$ is generically bigger for physical 6d CFTs, e.g. the minimal interacting CFT is the $A_1$ theory with $c=25$ \cite{Beem:2015aoa}, we plot the entire allowed range of $c$. In Figure \ref{6dbound} we show upper bounds on the OPE coefficients for the protected $\mathcal{B}[0,2]_\ell$ multiplet for the lowest spins $\ell=1,3$. While we cannot compute lower bounds as in 3d, because this multiplet is not separated from the continuum of long multiplets, the upper bound in this case is now nontrivial. We also show upper bounds on the lowest dimension scaling dimension $\Delta_{0,\ell}$ of the long multiplet with spin $\ell=2,4$. We compare these bounds to the 1-loop data for the pure AdS$_7$ theory as given in \eqref{6dresults}, as well as for the $AdS_7\times S^4$ and $AdS_7\times S^4/\mathbb{Z}_2$ theories \cite{Alday:2020tgi}, which we review in Appendix \ref{sec:tree}. Again, we find that the pure AdS$_7$ 1-loop correction noticeably saturates the numerical bounds relative to the tree, $AdS_7\times S^4$, or $AdS_7\times S^4/\mathbb{Z}_2$ expressions. We also computed an upper bound on $c$ (i.e. a lower bound on the stress tensor OPE coefficient), which applies to any interacting 6d $(2,0)$ CFT, and got
\es{cbound}{
c\geq21.6441\,,
}
which is weaker than the bound $c\gtrsim25$ conjectuerd in \cite{Beem:2015aoa}. This latter bound was found by extrapolating bounds computed at lower values of $\Lambda$ to $\Lambda\to\infty$, and was used as evidence that these general bootstrap bounds were saturated by the physical $A_1$ theory with $c=25$. We use a different definition of $\Lambda$ than \cite{Beem:2015aoa} \footnote{Compare our definition in 6.13 of \cite{Chester:2014fya}, to their definition in 5.9 of \cite{Beem:2015aoa}. We thank Balt van Rees for pointing this out.}, so it is hard check their conjectured extrapolation against our bound, but since in 3d \footnote{For the 3d $\mathcal{N}=8$ stress tensor bootstrap, a kink was found at $c\approx22.2735$ even at the high value of $\Lambda=43$ \cite{Agmon:2019imm}, which is close but different from the lowest known interacting theory of $U(2)_{2}\times U(1)_{-1}$ ABJ with $c\approx21.3333$. The value of $c$ at this kink was also shown to be  the lowest value allowed by a mixed correlator bootstrap that kinematically ruled out $U(2)_{2}\times U(1)_{-1}$ ABJ, which strongly suggests that even at infinite $\Lambda$ the kink will not correspond to this theory.} and 4d \footnote{For the 4d stress tensor bootstrap for $c=3/4$, corresponding to the lowest interacting $SU(2)$ SYM theory, it was shown in \cite{Chester:2021aun} that bounds obtained after imposing localization constraints for this theory strictly rule out the more general bounds like those in this paper.} we know that the general bounds are not saturated by the string/M-theory theory duals for the smallest such values of $c$, it seems likely that this general 6d bound is also not saturated by the $A_1$ theory even at $\Lambda\to\infty$.

\section{Discussion}\label{sec:discussion}

Our results show that pure AdS$_{d+1}$ maximal supergravity saturates the most general non-perturbative bootstrap bounds in the large $c$ regime, while CFTs with string/M-theory duals lie in the allowed region. This suggests that to study the latter theories, one needs to disallow the existence of the pure AdS$_{d+1}$ theory by either looking at mixed correlator with other single trace operators \cite{Agmon:2019imm,Bissi:2020jve}, or imposing theory specific constraints like supersymmetric localization \cite{Pestun:2016zxk}. Indeed, in 3d one can strengthen these general bootstrap bounds by inputting the OPE coefficients of the $(B,2)$ and $(B,+)$ multiplets for the $U(N)_k\times U(N)_{-k}$ ABJM theory for $k=1,2$, as computed to all orders in $1/N$ using localization in \cite{Agmon:2017xes}, in which case the 1-loop data for the dual $AdS_4\times S^7/\mathbb{Z}_k$ theories then saturates the bounds \cite{Alday:2021ymb,Alday:2022rly}. In 4d, one can input the two localization inputs for $SU(N)$ SYM derived in \cite{Binder:2019jwn,Chester:2020dja}, which are a function of the complexified coupling $\tau$, in which case the the bounds in \cite{Chester:2021aun} match 4-loop weak coupling results \cite{Fleury:2019ydf} in the appropriate regime, and exclude the general bootstrap bounds shown here for all $\tau$. In 6d there is no localization, but for correlators of  single trace operators other than the stress tensor one can input nontrivial OPE coefficients given by the protected 2d chiral algebra \cite{Beem:2014kka,Chester:2018dga} for the $A_{N-1}$ or $D_N$ theories

We can also use the general bootstrap bounds themselves to further study the pure AdS$_{d+1}$ theory, assuming it continues to saturate the bounds to higher order in $1/c$. In particular, by applying a fit to the large $c$ regime of the numerical bounds, one could read off higher derivative corrections to supergravity such as the $\mathcal{G}^{R^4}$ term discussed in the introduction, to help determine a putative UV completion. Since $\mathcal{G}^{R^4}$ occurs at the same order as higher loop corrections in some cases, e.g. $c^{-3}$ for pure AdS$_5$ \eqref{G}, it will be necessary to compute these higher loops, as was recently recently done for the 2-loop correction on $AdS_5\times S^5$ \cite{Huang:2021xws,Drummond:2022dxw}. The pure AdS$_{d+1}$ case should be much easier due to the lack of mixing, and so could even guide the calculation in the more physical cases with compact factors. More ambitiously, we can non-perturbatively define the pure AdS$_{d+1}$ theory as whatever saturates the bootstrap bounds at finite $c$; it would be fascinating to find independent evidence for or against the existence of such a theory.

Finally, we can ask what theory saturates the stress tensor correlator bootstrap bound with less than maximal supersymmetry. In 3d, the $\mathcal{N}=6$ bootstrap bounds were found in \cite{Binder:2020ckj,Binder:2021cif} to be saturated by $U(1)_{2N}\times U(1+N)_{-2N}$ ABJ theory \cite{Aharony:2008gk} for all $N$, which has a vector-like large $N$ limit dual to supersymmetric higher spin gravity \cite{Chang:2012kt,Aharony:2020omh,Aharony:2021ovo}. With no supersymmetry, it was observed in \cite{El-Showk:2014dwa,ElShowk:2012ht,Kos:2013tga} that critical $O(N)$ vector models saturate the bound on $c$ \footnote{See \cite{Chester:2015lej,Chester:2015qca} for similar results on $\mathcal{N}=2$ critical $O(N)$ vector models.}, so it is likely that the 3d stress tensor correlator bounds in general are saturated by interacting vector model CFTs. In higher dimensions, however, there are no interacting unitary vector models \footnote{The critical $O(N)$ vector model can be defined also in $4<d<6$ \cite{Fei:2014yja}, but it is non-unitary \cite{Giombi:2019upv}. Nonetheless, it can be non-rigorously bootstrapped with some success \cite{Chester:2014gqa,Li:2016wdp,Nakayama:2014yia}.}, so it is possible that the most general non-supersymmetric stress tensor bounds could be saturated by pure AdS$_{d+1}$ Einstein gravity with $d>3$. It would be fascinating to check this by generalizing the non-supersymmetric stress tensor bootstrap in 3d \cite{Dymarsky:2017yzx} to higher $d$. If such non-supersymmetric pure AdS$_{d+1}$ theories exist for any $d$, then they suggest that unitary interacting CFTs can be constructed for any $d$, unlike supersymmetric CFTs which only exist for $d\leq6$.

\section*{Acknowledgments}

We thank Anatoly Dymarsky, Balt van Rees, Joao Penedones, and Leonardo Rastelli for useful conversations, Himanshu Raj for collaboration on related projects, and Ofer Aharony for reviewing the manuscript. We also thank the organizers of the 2022 Bootstrap conference in Porto, during which this project was completed. SMC is supported by the Weizmann Senior Postdoctoral Fellowship. The work of LFA is supported by the European Research Council (ERC) under the European Union's Horizon
2020 research and innovation programme (grant agreement No 787185).  LFA is also supported in part by the STFC grant ST/T000864/1. The authors would like to acknowledge the use of the WEXAC cluster in carrying out this work.

\appendix

\section{Multiplets}\label{sec:multiplets}

\begin{table}
\centering
\begin{tabular}{|c|c|r|c|c|}
\hline
Type    & $(\Delta,\ell)$     & ${SO}(8)_R$ irrep  &spin $\ell$  &$1/c$ exact \\
\hline
$(B,+)$ &  $(2,0)$         & ${\bf 294}_c = [0040]$& $0$ & no \\ 
$(B,2)$ &  $(2,0)$         & ${\bf 300} = [0200]$& $0$ & no \\
$(B,+)$ &  $(1,0)$         & ${\bf 35}_c = [0020]$ & $0$& yes  \\
$(A,+)$ &  $(\ell+2,\ell)$       & ${\bf 35}_c = [0020]$ &even & no \\
$(A,2)$ &  $(\ell+2,\ell)$       & ${\bf 28} = [0100]$ & odd & no \\
Long &  $\Delta> \ell+1$ & ${\bf 1} = [0000]$ & even& no \\
$\text{Id}$ &  $(0,0)$ & ${\bf 1} = [0000]$ & even& N/A \\
\hline
\end{tabular}
\caption{The possible superconformal multiplets in the $S\times S$ OPE for 3d $\mathcal{N}=8$ CFTs.  The quantum numbers are those of the superconformal primary in each multiplet.}
\label{opemult3d}
\end{table} 

In this appendix we review the supermultiplets that appear in the OPE $S\times S$ for $d=3,4,6$ interacting theories. In 3d, $S$ is a $(B,+)$ type multiplet that transforms in the $[0020]$ of $SO(8)_R$, and we show the possible multiplets in Table \ref{opemult3d}. In this case, none of the protected multiplets are $1/c$ exact except trivially the stress tensor multiplet itself. 

\begin{table}
\centering
\begin{tabular}{|c|c|r|c|c|}
\hline
Type    & $(\Delta,\ell)$     & ${SU}(4)_R$ irrep  &spin $\ell$  &$1/c$ exact \\
\hline
$\mathcal{B}$ &  $(2,0)$         & ${\bf 20'} = [020]$ & $0$& yes  \\
$\mathcal{B}$ &  $(4,0)$         & ${\bf 105} = [040]$ & $0$& yes  \\
$\mathcal{B}$ &  $(4,0)$         & ${\bf 84} = [202]$ & $0$& yes  \\
$\mathcal{C}$ &  $(\ell+4,\ell)$       & ${\bf 20'} = [020]$ & even & yes \\
$\mathcal{C}$ &  $(\ell+4,\ell)$       & ${\bf 15} = [101]$ & odd & yes \\
Long &  $\Delta> \ell+2$ & ${\bf 1} = [000]$ & even& no \\
$\text{Id}$ &  $(0,0)$ & ${\bf 1} = [000]$ & even& N/A \\
\hline
\end{tabular}
\caption{The possible superconformal multiplets in the $S\times S$ OPE for 4d $\mathcal{N}=4$ CFTs.  The quantum numbers are those of the superconformal primary in each multiplet, and for familiarity we use $SU(4)$ conventions for the Dynkin labels.}
\label{opemult4d}
\end{table}

In 4d, $S$ is a $\mathcal{B}$ type multiplet in the $[020]$ of $SU(4)_R$, and we show the possible multiplets in Table \ref{opemult4d}. Here, there are no non-trivial protected multiplets.

\begin{table}
\centering
\begin{tabular}{|c|c|r|c|c|}
\hline
Type    & $(\Delta,\ell)$     & ${SO}(5)_R$ irrep  &spin $\ell$  &$1/c$ exact \\
\hline
$\mathcal{D}$ &  $(4,0)$         & ${\bf 14} = [20]$ & $0$& yes  \\
$\mathcal{D}$ &  $(8,0)$         & ${\bf 35'} = [04]$ & $0$& no  \\
$\mathcal{D}$ &  $(8,0)$         & ${\bf 55} = [40]$ & $0$& yes  \\
$\mathcal{B}$ &  $(\ell+8,\ell)$       & ${\bf 14} = [20]$ &even & yes \\
$\mathcal{B}$ &  $(\ell+8,\ell)$       & ${\bf 10} = [02]$ & odd & no \\
Long &  $\Delta> \ell+6$ & ${\bf 1} = [00]$ & even& no \\
$\text{Id}$ &  $(0,0)$ & ${\bf 1} = [00]$ & even& N/A \\
\hline
\end{tabular}
\caption{The possible superconformal multiplets in the $S\times S$ OPE for 6d $(2,0)$ CFTs.  The quantum numbers are those of the superconformal primary in each multiplet.}
\label{opemult6d}
\end{table} 

In 6d, $S$ is a $\mathcal{D}$ type multiplet in the $[20]$ of $SO(5)_R$, and we show the possible multiplets in Table \ref{opemult6d}. Here, the non-trivial protected multiplets are $\mathcal{D}[04]$ and $\mathcal{B}[02]_\ell$ with odd $\ell$, which are identical to the long multiplets at their unitarity value $\ell=6$. 

\section{CFT data}\label{sec:tree}

In this appendix, we collect previous results for 1-loop CFT data in $d=3,4,6$ for string/M-theory duals, which we will use in the main text. In 3d, 
 the 1-loop corrections were computed for $U(N)_k\times U(N)_{-k}$ ABJM dual to $AdS_4\times S^7/\mathbb{Z}_k$ for $k=1,2$ in \cite{Alday:2021ymb,Alday:2022rly} to get for the $k=1$ theory
\es{3dresults2}{
&AdS_4\times S^7:\qquad\qquad\qquad\quad\;\;\gamma^{R|R}_{0,2}=-39254.4\,,\\
&(\lambda^{R|R}_{(A,+)_0})^2=513.49\,,\qquad (\lambda^{R|R}_{(A,2)_1})^2=4170.0\,,\\
}
and for the $k=2$ theory
\es{3dresults3}{
&AdS_4\times S^7/\mathbb{Z}_2:\qquad\qquad\quad\;\;\;\,\gamma^{R|R}_{0,2}=-16740.9\,,\\
&(\lambda^{R|R}_{(A,+)_0})^2=285.32\,,\qquad (\lambda^{R|R}_{(A,2)_1})^2=2239.9\,.\\
}

In 4d, 
the 1-loop corrections were computed for $\mathcal{N}=4$ SYM with $SU(N)$ \cite{Alday:2017xua,Aprile:2017bgs} and $SO(N)$ \cite{Alday:2021vfb} gauge group dual to $AdS_5\times S^5$ and $AdS_5\times S^5/\mathbb{Z}_2$, respectively, to get
\es{4dresults2}{
&AdS_5\times S^5:\qquad\quad\;\;\gamma^{R|R}_{0,2}=-2.5625\,,\\
&AdS_5\times S^5/\mathbb{Z}_2:\qquad\gamma^{R|R}_{0,2}=-0.88851\,.\\
}

In 6d, 
the 1-loop corrections were computed for $A_{N-1}$ and $D_N$ CFTs dual to $AdS_7\times S^4$ and $AdS_7\times S^4/\mathbb{Z}_2$, respectively, to get for the former theory
\es{6dresults2}{
&AdS_7\times S^4:\\
&\quad\;\;\,\gamma^{R|R}_{0,2}=-1171.1\,,\qquad\quad\;\; \;\;\gamma^{R|R}_{0,4}=-25.414\,,\\
&(\lambda^{R|R}_{\mathcal{B}[02]_1})^2=-12.388\,,\qquad (\lambda^{R|R}_{\mathcal{B}[02]_3})^2=-0.18697\,,
}
and for the latter theory
\es{6dresults3}{
&AdS_7\times S^4/\mathbb{Z}_2:\\
&\quad\;\;\;\,\gamma^{R|R}_{0,2}=-644.25\,,\qquad\quad\;\; \,\,\gamma^{R|R}_{0,4}=-18.918\,,\\
&(\lambda^{R|R}_{\mathcal{B}[02]_1})^2=-7.6294\,,\qquad (\lambda^{R|R}_{\mathcal{B}[02]_3})^2=-0.15983\,.
}

\section{Inversion formulae}\label{sec:inverse}

In this appendix we collect the inversion formulae from \cite{Alday:2021ymb,Alday:2020tgi} that we apply to the DDs computed for the 1-loop pure AdS$_{d+1}$ correlator for $d=3,6$ to get the CFT data reported in the main text. Recall that for 4d, the pure AdS$_5$ results are already available from \cite{Alday:2017xua}. For 3d, the $(A,+)_\ell$ formula was given in \eqref{ApInversion}, while the $(A,2)_\ell$ result can be extracted from the formula
\es{0200}{
&\frac{12 (\ell+1)^2 (\ell+2)^2}{(2 \ell+1) (2 \ell+3)^2 (2 \ell+5)} \lambda^2_{(A,2)_{\ell-1}}\\
&+
\frac{2 (\ell+2) (\ell+3)}{(2 \ell+3) (2 \ell+7)}\lambda^2_{(A,+)_\ell}+\frac{3}{4}\lambda^2_{(A,2)_{\ell+1}} = \\
&\frac{12 (2 \ell+5) \Gamma (\ell+3)^4}{\Gamma \left(\ell+\frac{5}{2}\right)^2
   \Gamma \left(\ell+\frac{7}{2}\right)^2}\\
   &\times \int_0^1 \frac{d \bar z}{\bar z}    g_{\ell+4,\ell+2}(\bar z) \text{dDisc}[ {\cal G}^{[0200]}(z\bar z,1-\bar z,)\vert_z ]  \,,
   }
after plugging in the results for $\lambda^2_{(A,+)_\ell}$, and using the lightcone block with normalization
\es{lightconeBlock}{
&g_{\Delta,\ell}(1-V)=\frac{\Gamma(\ell+1/2)}{4^\Delta\sqrt{\pi}\ell!}(1-V)^\ell \\
&\quad\times{}_2F_1\left(\frac{\Delta+\ell}{2},\frac{\Delta+\ell}{2},\Delta+\ell,1-V\right)\,.\\
}
The $(B,+)$ and $(B,2)$ OPE coefficients then correspond to the values
\es{B2Bp}{
 \lambda^2_{(B,2)}= \lambda^2_{(A,2)_{-1}}\,,\qquad  \lambda^2_{(B,+)}= \lambda^2_{(A,+)_{-2}}\,. 
}
We can extract the anomalous dimension from the formula
\es{anom1}{
\gamma^{R|R}_{0,\ell} =& \frac{1}{(\lambda^{(0)}_{0,\ell})^2 }\Big(4 R^{[0040]}_{1,R|R}( \ell) + \frac12 \partial_{\ell} \big[(\lambda^{(0)}_{0,\ell})^2 ( \gamma^{R}_{0,\ell} )^2\big] \\
&- (\lambda^{R}_{0,\ell})^2  \gamma^{R}_{0,\ell} \Big)\\
&+\left(A_{2\ell}-\frac{7}{12}\right)(\gamma^R_{0,\ell})^2\,,
}
where $A_n=\sum_{k=1}^n(-1)^{k+1}/k$ and $A_0=0$. The last term retains the dimension dependence of the conformal-block normalization and the superblock coefficient. The inversion integral is
\es{anom2}{
&R^{[0040]}_{1,R|R}(\ell) =\frac{512 (\ell+1) (\ell+2) (2 \ell+3) \Gamma (\ell+1)^4}{\Gamma
   \left(\ell+\frac{1}{2}\right)^2 \Gamma \left(\ell+\frac{5}{2}\right)^2}\\
   &\times\int_0^1 {d \bar z}{\bar z}  g_{\ell+6,\ell}(\bar z) \text{dDisc}\left. {\cal G}_{R|R}^{[0040]}(z\bar z,1-\bar z) \right|_{z^3 \log z} \,,
}
and the tree and GFFT formula needed above take a more complicated form that we give in the attached \texttt{Mathematica} file.

For 6d, it is convenient to solve the superconformal Ward identities by writing $\mathcal{G}(U,V;\sigma,\tau)$ in \eqref{4point} as
\es{6dred}{
\mathcal{G}(U,V;\sigma,\tau)=\mathcal{F}(U,V;\sigma,\tau)+\Upsilon\circ \mathcal{H}(U,V)\,,
}
where $\mathcal{F}$ is the free theory correlator,  $\Upsilon$ is a complicated differential operator defined in \cite{Dolan:2004mu}, and $\mathcal{H}(U,V)$ is an R-symmetry singlet called the reduced correlator. In terms of this reduced correlator, we can extract the 1-loop OPE coefficient as
\es{inversion}{
&\lambda^2_{\mathcal{B}[02]_\ell}  =  -\frac{\pi (\ell+1) (\ell+4) \Gamma (\ell+5) \Gamma (\ell+7)}{ 2^{4 \ell+19} (\ell+2) \Gamma \left(\ell+\frac{11}{2}\right) \Gamma \left(\ell+\frac{13}{2}\right)} \\
&\times\int_0^1 d\bar z  \bar z^{4} g_{\ell+11,\ell+1}^{-2,0}(\bar z)\text{dDisc} \left.\mathcal{H}(z\bar z,1-\bar z) \right|_{z^0}\,,
}
where we define the mixed lightcone block in the 6d normalization as
\es{lightblock2}{
&g_{\Delta,\ell}^{-2,0}(1-V)=(V-1)^\ell \\
&\qquad\times{}_2F_1\left(\frac{\Delta+\ell+2}{2},\frac{\Delta+\ell}{2},\Delta+\ell,1-V\right)\,.\\
}
The $\mathcal{D}[04]$ OPE coefficient then corresponds to the limit
\es{D04}{
 \lambda^2_{\mathcal{D}[04]}= \lim_{\ell\to-1} (\ell+1)\lambda^2_{\mathcal{B}[02]_{\ell}}\,.
 }
We can extract the anomalous dimension from the formula
\es{hatNohat}{
&\gamma^{R|R}_{0,\ell}=\hat\gamma^{R|R}_{0,\ell}+\gamma^\text{extra}_\ell\,,\\
&\gamma^\text{extra}_\ell\equiv -\frac{298598400 (2 \ell+11) \left(\ell^2+11 \ell+14\right)}{(\ell+1)^3 (\ell+2)^3 (\ell+9)^3 (\ell+10)^3}\,,
}
where we compute
\es{hat}{
&\hat \gamma^{R|R}_{0,\ell} = -\frac{\sqrt{\pi}45\ 2^{-7-2\ell} \Gamma (\ell+5)}{ (\ell+1) (\ell+2) (\ell+9) (\ell+10) \Gamma \left(\ell+\frac{13}{2}\right)} \\
&\times \int_0^1 d\bar z \bar z^{4} g_{\ell+11,\ell+1}^{-2,0}(\bar z) \Big[\text{dDisc}\left.\mathcal{H}(z\bar z,1-\bar z) \right|_{\log z}\\
&- \frac{3456 \left(397 \bar z^3-2910 \bar z^2+5730 \bar z-3305\right)}{\bar z^5 (1-\bar z)^{-1}}\Big]\,.
}

\onecolumngrid
\vspace{1in}
\twocolumngrid

\bibliographystyle{ssg}
\bibliography{sat}

\end{document}